\documentclass[
reprint,
amsmath,
amssymb,
aps,
prb,
superscriptaddress
]{revtex4-2}

\usepackage{color}
\usepackage{multirow}
\usepackage{graphicx}
\usepackage{dcolumn}
\usepackage{bm}
\usepackage{hyperref}
\hypersetup{
    colorlinks = true,
    linkcolor = blue,
    citecolor = blue,
    urlcolor = blue,
    pdfborder = {0 0 0},
    pdfusetitle = true
}

\begin{document}

\title{Small-supercell and Small-dataset Training Strategy of Machine Learning Interatomic Potentials for Point Defects
}%

\author{Zhenxing Dai}
    \affiliation{
    Key Laboratory of Computational Physical Sciences (MOE), Fudan University, Shanghai 200433, China
    }
    \affiliation{Department of Physics, Fudan University, Shanghai 200433, China}
\author{Mingjue Ni}
    \affiliation{
    Key Laboratory of Computational Physical Sciences (MOE), Fudan University, Shanghai 200433, China
    }
    \affiliation{Department of Physics, Fudan University, Shanghai 200433, China}
\author{Xinpeng Li}
    \affiliation{
    Key Laboratory of Computational Physical Sciences (MOE), Fudan University, Shanghai 200433, China
    }
    \affiliation{Department of Physics, Fudan University, Shanghai 200433, China}
\author{Menglin Huang}
    \email{Contact author: menglinhuang@fudan.edu.cn}
    \affiliation{
    Key Laboratory of Computational Physical Sciences (MOE), Fudan University, Shanghai 200433, China
    }
    \affiliation{College of Integrated Circuits and Micro-Nano Electronics, Fudan University, Shanghai 200433, China}
\author{Anderson Janotti}
    \affiliation{
    Department of Materials Science and Engineering, University of Delaware, Newark, Delaware, 19716, USA
    }
\author{Shiyou Chen}
    \email{Contact author: chensy@fudan.edu.cn}
    \affiliation{
    Key Laboratory of Computational Physical Sciences (MOE), Fudan University, Shanghai 200433, China
    }
    \affiliation{College of Integrated Circuits and Micro-Nano Electronics, Fudan University, Shanghai 200433, China}

\begin{abstract}

Machine learning interatomic potentials (MLIPs) can treat large-scale material systems with near first-principles accuracy and have been widely used to accelerate point-defect simulations. However, the training of MLIPs usually relies on large amounts of DFT data. This issue is particularly pronounced for charged defects, for which DFT calculations of large supercells are required to avoid long-range Coulomb interactions and finite-size effects, making the construction of datasets computationally expensive. In this work, we propose an efficient MLIP training scheme for neutral and lowly charged point defects based on small supercells (less than 100 atoms) and limited number of DFT calculations. The scheme requires only four DFT structural relaxations to construct the training dataset and the trained MLIPs dedicated for the defect can predict the defect formation energies in larger supercells (over 200 atoms) with small errors (mostly smaller than 0.3 eV). Using defects in GaN, SiO$_2$, and $\mathrm{Cu}_2\mathrm{ZnSnS}_4$ as representative examples, we evaluate the extrapolation capability of this scheme for predicting defect total energies and structural relaxations across different supercell sizes. The results show that an MLIP trained only on single small defect supercell produces severe errors when treating large supercells. Incorporating defect data from multiple supercell sizes improves the predictive performance of the MLIP, while adding pristine defect-free bulk supercells further enhances the accuracy. These results provide practical guidance for training MLIP models of defect systems with low computational costs.

\end{abstract}

\maketitle

\section{Introduction}

Point defects play a central role in determining the properties of semiconductor materials and the performance of electronic and optoelectronic devices~\cite{ganose2022defect,kimoto2020defect,park2018point}. Because their microscopic structures are often difficult to characterize directly in experiments, first-principles calculations based on density functional theory (DFT) have long served as an important theoretical tool for defect studies. However, DFT calculations of point defects generally need to account for the defect species, local atomic configurations, and charge states simultaneously. In particular, charged defects are strongly affected by finite-size effects, so large supercells containing hundreds of atoms and appropriate electrostatic corrections are often needed to obtain converged formation energies and charge transition levels. These requirements substantially increase the computational cost and limit the efficiency of systematic defect calculations.

Machine-learning interatomic potentials (MLIPs)~\cite{zhang2018deep,batatia2022mace,deng2023chgnet,batzner20223,musaelian2023learning} have recently become widely used for accelerating atomistic simulations of materials. By learning potential energy surfaces from first-principles data, MLIPs can predict energies, atomic forces, and stresses at much lower computational cost, while retaining near first-principles accuracy in many bulk systems. In defect studies, MLIPs make it possible to access larger length scales and longer time scales than direct DFT calculations, enabling applications such as the search for metastable defect configurations~\cite{morrow2024understanding,jiang2024machine,li2025machine}, the simulation of defect formation and migration at finite temperatures~\cite{wu2023oxygen,chen2025simulating}, and the modeling of irradiation-induced damage processes~\cite{song2023neural,chen2024deep}. These applications demonstrate the potential of MLIPs for large-scale defect problems that are difficult to address using DFT alone.

Despite this progress, constructing reliable MLIPs for defect systems remains computationally demanding. Most current foundation models are pretrained on existing databases of perfect crystalline materials, whereas high-quality defect datasets are still much less represented. As a result, pretrained models can often describe ideal crystals with reasonable accuracy, but their errors are typically much larger for defect  systems~\cite{wang2024perovsdopants,wang2025evaluating}, which can hinder the prediction of ground-state and metastable defect configurations~\cite{wang2026multi}. In many defect-related MLIP studies, new DFT calculations must therefore be carried out for each target defect system, and the resulting training datasets may contain thousands or even tens of thousands of DFT data points~\cite{song2023neural,yang2025impact,dou2025machine,morrow2024understanding}. This process is time-consuming and can substantially reduce the practical efficiency advantage of MLIPs. Developing training schemes that can work with limited DFT data is therefore essential for extending MLIP-based simulations to broader defect studies.

The data-efficiency problem is particularly important for charged defects. Under periodic boundary conditions, a charged defect interacts electrostatically with its periodic images, and the compensating background charge introduced to maintain charge neutrality also produces spurious electrostatic contributions. In conventional DFT defect calculations, these finite-size effects are treated using correction schemes ~\cite{lany2008assessment,freysoldt2009fully,kumagai2014electrostatics}. Standard MLIPs, however, describe atomic structures mainly through elemental species and local environments within a finite cutoff radius, and therefore cannot explicitly represent long-range Coulomb interactions beyond this cutoff. Consequently, the model may learn not only the local bonding environment, but also finite-size errors contained in the DFT reference data. Since these contributions depend on the supercell size and boundary conditions rather than only on the local structure, they can limit the ability of standard MLIPs to extrapolate charged-defect energies and relaxed structures from small to large supercells. In addition, most MLIPs do not include the global charge state as an explicit input, so mixing different charge states in one model may map configurations belonging to different potential energy surfaces onto the same descriptor space~\cite{wang2025evaluating}.

Current efforts to model charged defects with MLIPs generally follow two directions. One direction is to introduce additional physical information into the model, for example by explicitly treating long-range electrostatics through the Deep Potential Long Range framework~\cite{zhang2022deep}, Ewald summation~\cite{cheng2025latent}, or reciprocal-space information~\cite{guo2026capturing}, or by incorporating the global charge state as an input variable~\cite{shimizu2022using,dou2025machine}. These approaches are physically appealing, but their applications to charged point defects in semiconductors are still developing. The other, more commonly used approach is to train separate MLIPs for defects in different charge states, thereby avoiding the direct mixing of different charge-state potential energy surfaces~\cite{mosquera2025point,chen2025simulating,rahman2026defect,zhou2025one}. In this work, we use the term standard MLIPs to refer to MLIP models that do not explicitly include long-range electrostatic interactions, global charge information, or other additional physical descriptors. However, separate training for each charge state does not by itself resolve the finite-size errors present in charged-defect DFT data, including large charged-defect supercells would also substantially increase the cost of dataset construction. For such models, a central practical question is therefore whether the composition of a small DFT training dataset can be designed so that the resulting MLIP remains reliable when applied to larger charged-defect supercells.

In this work, we propose an efficient MLIP training scheme for charged point defects based on small supercells and limited DFT data. The scheme constructs the training dataset from the relaxation trajectories of only four DFT structural optimizations and is designed to predict charged-defect properties in larger supercells. We assess this strategy using three representative semiconductor defect systems with different bonding environments and chemical complexity: $V_{\mathrm{N}}$ in GaN, $V_{\mathrm{O}}$ in SiO$_2$, and $\mathrm{Cu}_{\mathrm{Zn}}$ in $\mathrm{Cu}_2\mathrm{ZnSnS}_4$. For these systems, we evaluate the ability of standard MLIPs trained with limited small-supercell data to extrapolate defect total energies and structural relaxations across different supercell sizes. Our results show that the data composition is critical for charged-defect MLIPs. A model trained only on charged-defect configurations from a single small supercell size produces severe errors when applied to larger defect supercells. Introducing charged-defect DFT data from multiple small supercell sizes substantially reduces both energy and structural errors. Further adding neutral pristine bulk configurations improves the prediction accuracy and structural stability for defects in low charge states. Nevertheless, for highly charged defects, larger charged-defect supercells still need to be included in the training set to obtain reliable extrapolation. These findings clarify the practical limits of standard MLIPs for charged-defect calculations and provide guidance for constructing reliable charged-defect training datasets from limited DFT data within a general MLIP framework.

\section{Methods}

\subsection{First-principles calculations}

All first-principles calculations in this work were performed using the VASP package~\cite{kresse1996efficient}. The electronic exchange-correlation energy was described by the Perdew--Burke--Ernzerhof (PBE) functional within the generalized gradient approximation (GGA)~\cite{perdew1996generalized,perdew1997generalized}. Spin polarization was considered for all systems. The interaction between ionic cores and valence electrons was treated using the projector augmented-wave (PAW) method~\cite{kresse1999ultrasoft}. The plane-wave basis-set cutoff energy was chosen according to the material system, with values of 400~eV for GaN and SiO$_2$ and 295~eV for $\mathrm{Cu}_2\mathrm{ZnSnS}_4$. During structural relaxation, the energy convergence criterion for electronic self-consistent iterations was set to $10^{-5}$~eV, and the maximum force convergence criterion for ionic relaxation was set to 0.02~eV/\AA.

To maintain comparable reciprocal-space sampling densities among supercells of different sizes, the $\Gamma$-centered $k$-point meshes were adjusted according to the changes in the supercell lattice constants for each system. As the supercell size increased, the number of $k$ points along each direction was reduced in accordance with the increase in the corresponding real-space lattice length. The specific lattice dimensions, numbers of atoms, and corresponding $k$-point settings for each system are given in Table~\ref{tab:kpoints}.

\vspace{5pt}
\begin{table*}[!htbp]
\centering
\setlength{\tabcolsep}{12pt}
\renewcommand{\arraystretch}{1.3}
\caption{ Lattice constants, total number of atoms, and corresponding $k$-point meshes for GaN, $\mathrm{Cu}_2\mathrm{ZnSnS}_4$, and SiO$_2$ supercells.}
\label{tab:kpoints}
\begin{tabular}{c|c|c|c}
\hline\hline
System & $a \times b \times c$ (\AA$^3$) & Total atoms & $k$-point mesh \\
\hline
\multirow{6}{*}{GaN} 
 & 6.45 $\times$ 5.59 $\times$ 5.25    & 15  & $5 \times 6 \times 6$ \\
 & 6.45 $\times$ 5.59 $\times$ 10.50   & 31  & $5 \times 6 \times 3$ \\
 & 6.45 $\times$ 11.18 $\times$ 10.50  & 63  & $5 \times 3 \times 3$ \\
 & 6.45 $\times$ 11.18 $\times$ 15.75  & 95  & $5 \times 3 \times 2$ \\
 & 12.91 $\times$ 11.18 $\times$ 10.50 & 127 & $3 \times 3 \times 3$ \\
 & 12.91 $\times$ 11.18 $\times$ 21.00 & 255 & $3 \times 3 \times 2$ \\
\hline
\multirow{5}{*}{$\mathrm{Cu}_2\mathrm{ZnSnS}_4$}    
 & 5.47 $\times$ 5.47 $\times$ 10.92   & 16  & $6 \times 6 \times 3$ \\
 & 10.94 $\times$ 5.47 $\times$ 10.92  & 32  & $3 \times 6 \times 3$ \\
 & 10.94 $\times$ 10.94 $\times$ 10.92 & 64  & $3 \times 3 \times 3$ \\
 & 10.94 $\times$ 10.94 $\times$ 21.83 & 128 & $3 \times 3 \times 2$ \\
 & 16.42 $\times$ 16.42 $\times$ 21.83 & 288 & $2 \times 2 \times 2$ \\
\hline
\multirow{4}{*}{SiO$_2$} 
 & 8.51 $\times$ 4.91 $\times$ 5.43    & 17  & $4 \times 6 \times 6$ \\
 & 8.51 $\times$ 9.83 $\times$ 5.43    & 35  & $4 \times 3 \times 6$ \\
 & 8.51 $\times$ 9.83 $\times$ 10.86   & 71  & $4 \times 3 \times 3$ \\
 & 17.03 $\times$ 14.74 $\times$ 16.29 & 323 & $2 \times 2 \times 2$ \\
\hline\hline
\end{tabular}
\end{table*}

\subsection{Generation of training data}

All defect configurations in the training sets were extracted from DFT structural relaxation trajectories. The initial geometries for these trajectories were explicitly chosen based on the specific defect type and charge state. For neutral systems, structural relaxations were initiated from unrelaxed configurations, which were generated by directly introducing a vacancy or substitutional defect into the pristine lattice. For charged systems, however, the initialization strategy varied. Charged vacancy defects, such as $V_{\mathrm{N}}^{+}$ in GaN and $V_{\mathrm{O}}^{+}$ in SiO$2$, were relaxed starting from the fully optimized equilibrium structures of their corresponding neutral states. Conversely, for the substitutional defect $\mathrm{Cu}_{\mathrm{Zn}}$ in $\mathrm{Cu}_2\mathrm{ZnSnS}_4$, relaxations proceeded directly from the unrelaxed substituted configuration. This tailored approach accounts for the varying degrees of local structural reconstruction induced by charge-state transitions. Vacancy defects typically drive pronounced atomic displacements and local rearrangements; therefore, initiating the relaxation from the neutral defect's equilibrium geometry better mimics the structural evolution that occurs following carrier trapping or emission. In contrast, the relatively weak lattice distortions induced by substitutional defects justify the direct use of unrelaxed configurations as initial structures.

The training sets for neutral defect systems is constructed from relatively small defect supercells and their corresponding pristine bulk configurations. The model's capacity for accurate structural relaxation and energy prediction is subsequently evaluated by extrapolating to larger supercells. A comprehensive breakdown of the training-set compositions is provided in Table~\ref{tab:neutral_defects}.

\begin{table*}[!htbp]
\setlength{\tabcolsep}{2.5mm}
\renewcommand{\arraystretch}{1.3}
\centering
\caption{Training-set compositions for neutral defects.}
\label{tab:neutral_defects}
\begin{tabular}{lcll}
\hline\hline
\shortstack{System} & \shortstack{Training-set index} & \shortstack{Defect supercell size (atoms)}  & \shortstack{Configuration numbers (total)} \\
\hline
\multirow{4}{*}{$\mathrm{GaN}-V_{\mathrm{N}}$} 
& I & 15 & 12+1 (13)\\
& II & 15+31 & 12+14+2 (28) \\
& III & 15+31+63 & 12+14+12+3 (41)\\
& IV & 15+31+63+95 & 12+14+12+14+4 (56)\\
\hline
\multirow{3}{*}{SiO$_{2}-V_{\mathrm{O}}$} 
& I & 17 & 80+1 \\
& II & 17+35 & 80+61+2 (143) \\
& III & 17+35+71 & 80+61+55+3 (199) \\
\hline
\multirow{3}{*}{$\mathrm{Cu}_2\mathrm{ZnSnS}_4 - \mathrm{Cu}_\mathrm{Zn}$} 
& I & 16 & 8+1 \\
& II & 16+32 & 8+8+2 (18) \\
& III & 16+32+64 & 8+8+8+3 (27) \\
\hline\hline
\end{tabular}
\end{table*}

The charged defect systems considered in this work include $V_{\mathrm{N}}^{+}$, $V_{\mathrm{N}}^{2+}$, and $V_{\mathrm{N}}^{3+}$ in GaN, $V_{\mathrm{O}}^{+}$ and $V_{\mathrm{O}}^{2+}$ in SiO$_2$, and $\mathrm{Cu}_{\mathrm{Zn}}^{-}$ in $\mathrm{Cu}_2\mathrm{ZnSnS}_4$. Because charged defect systems involve long-range Coulomb interactions and the corresponding finite-size effects, two training schemes were constructed to compare the influence of training-set composition on the energy prediction results of standard MLIPs.

\begin{table*}[!htbp]
\setlength{\tabcolsep}{2mm}
\renewcommand{\arraystretch}{1.3}
\centering
\caption{Training sets for charged defects in the Defect-Only Scheme.}
\label{tab:MLIP_scheme1}
\begin{tabular}{lcll}
\hline\hline
Defect system 
& \shortstack{Training-set index} 
& \shortstack{Defect supercell sizes (atoms)} 
& \shortstack{Configuration numbers (total)} \\
\hline
\multirow{4}{*}{$\mathrm{GaN}-V_{\mathrm{N}}^{+}$} 
& I & 15 & 12 \\
& II & 15+31 & 12+14 (26) \\
& III & 15+31+63 & 12+14+14 (40) \\
& IV & 15+31+63+95 & 12+14+14+14 (54) \\
\hline
\multirow{3}{*}{SiO$_{2}-V_{\mathrm{O}}^{+}$} 
& I & 17 & 93 \\
& II & 17+35 & 93+46 (139) \\
& III & 17+35+71 & 93+46+44 (183) \\
\hline
\multirow{3}{*}{SiO$_{2}-V_{\mathrm{O}}^{2+}$} 
& I & 17 & 117 \\
& II & 17+35 & 117+99 (216) \\
& III & 17+35+71 & 117+99+140 (356) \\
\hline
\multirow{3}{*}{$\mathrm{Cu}_2\mathrm{ZnSnS}_4 - \mathrm{Cu}_\mathrm{Zn}^{-}$} 
& I & 16 & 10 \\
& II & 16+32 & 10+10 (20) \\
& III & 16+32+64 & 10+10+8 (28) \\
\hline\hline
\end{tabular}
\end{table*}

In the Defect-Only Scheme, the training set contains only configurations sampled from DFT structural relaxation trajectories of charged defects and adopts an incremental multi-size supercell setup. Training is first performed using data from a single small supercell containing fewer than 20 atoms, after which charged-defect configurations in larger supercells containing up to 95 atoms are gradually introduced. The detailed training-set configuration for this scheme is given in Table~\ref{tab:MLIP_scheme1}.

\vspace{10pt}
\begin{table*}[!htbp]
\setlength{\tabcolsep}{2mm}
\renewcommand{\arraystretch}{1.3}
\centering
\caption{Training sets for charged defects in the Defect+Host Scheme.}
\label{tab:MLIP_scheme2}
\begin{tabular}{lcll}
\hline\hline
Defect system 
& \shortstack{Training-set index} 
& \shortstack{Defect supercell sizes (atoms)} 
& \shortstack{Configuration numbers (total)} \\
\hline
\multirow{4}{*}{$\mathrm{GaN}-V_{\mathrm{N}}^{+}$} 
& I & 15 & 12+1 (13) \\
& II & 15+31 & 12+14+2 (28) \\
& III & 15+31+63 & 12+14+14+3 (43) \\
& IV & 15+31+63+95 & 12+14+14+14+4 (58) \\
\hline
\multirow{4}{*}{$\mathrm{GaN}-V_{\mathrm{N}}^{2+}$} 
& I & 15 & 18+1 (19) \\
& II & 15+31 & 18+21+2 (41) \\
& III & 15+31+63 & 18+21+24+3 (66) \\
& IV & 15+31+63+95 & 18+21+24+27+4 (94) \\
\hline
\multirow{4}{*}{$\mathrm{GaN}-V_{\mathrm{N}}^{3+}$} 
& I & 15 & 18+1 (19) \\
& II & 15+31 & 18+24+2 (44) \\
& III & 15+31+63 & 18+24+25+3 (70) \\
& IV & 15+31+63+95 & 18+24+25+28+4 (99) \\
\hline
\multirow{3}{*}{SiO$_{2}-V_{\mathrm{O}}^{+}$} 
& I & 17 & 93+1 (94) \\
& II & 17+35 & 93+46+2 (141) \\
& III & 17+35+71 & 93+46+44+3 (186) \\
\hline
\multirow{3}{*}{SiO$_{2}-V_{\mathrm{O}}^{2+}$} 
& I & 17 & 117+1 (118) \\
& II & 17+35 & 117+99+2 (218) \\
& III & 17+35+71 & 117+99+140+3 (359) \\
\hline
\multirow{3}{*}{$\mathrm{Cu}_2\mathrm{ZnSnS}_4 - \mathrm{Cu}_\mathrm{Zn}^{-}$} 
& I & 16 & 10+1 (11) \\
& II & 16+32 & 10+10+2 (22) \\
& III & 16+32+64 & 10+10+8+3 (31) \\
\hline\hline
\end{tabular}
\end{table*}

In the Defect+Host Scheme, while retaining charged-defect configurations of different sizes, the training set also includes size-matched neutral defect-free bulk supercell data. This scheme is used to examine whether neutral pristine-lattice data can provide additional constraints on bulk-like local atomic environments and further affect the accuracy of standard MLIPs in predicting the relaxed configurations and energies of charged defects. The detailed training-set configuration for this scheme is given in Table~\ref{tab:MLIP_scheme2}.

\subsection{Training and evaluation of the MLIP} 

To avoid the mapping conflicts between potential energy surfaces associated with different charge states, an independent MLIP was trained from scratch for each specific defect type and its corresponding charge state.

We use Allegro package~\cite{batzner20223,musaelian2023learning} to train the MLIP model. In the parameter settings of the Allegro, the cutoff radius for the local atomic environment, $r_{\mathrm{cut}}$, was set to 6~\AA. The maximum order of the spherical harmonic expansion was set to $l_{\mathrm{max}}=2$, and the full O(3) parity symmetry was retained. The model contained two interaction layers, and the environment embedding multiplicity was set to 16. The hidden-layer dimensions of the multilayer perceptrons (MLPs) used for the two-body latent embedding and feature extraction were set to [64, 64, 128, 128, 128] and [128, 128, 128], respectively. The loss function was defined as the sum of the mean squared errors (MSEs) of the atomic force error and the per-atom energy error, with both weights set to 1.0.

The model was trained using the Adam optimizer with an initial learning rate of 0.002. To improve training stability, a ReduceLROnPlateau learning-rate decay strategy was introduced; when the validation loss did not decrease for 20 consecutive epochs, the learning rate was reduced by a factor of 0.5. The maximum number of training epochs was set to 100000, and an early-stopping mechanism was adopted. Training was terminated if the validation loss did not improve for 50 consecutive epochs. Datasets were randomly divided into training and validation sets at a ratio of 8.5:1.5 for SiO$_2$ and $\mathrm{Cu}_2\mathrm{ZnSnS}_4$, 8:2 for GaN.

After training, the MLIP model was exported and integrated into the LAMMPS software package~\cite{thompson2022lammps} to evaluate its generalization and size-extrapolation capability in larger defect supercells. To ensure a consistent evaluation of the MLIP's performance in structural optimization and energy prediction, the test-set configurations were initialized using the exact same protocols employed during the training-set generation. Specifically, structural relaxations for neutral defects and the charged substitutional defect ($\mathrm{Cu}_{\mathrm{Zn}}^{-}$) started directly from unrelaxed configurations, which were generated by introducing the defect into the perfect host lattice. In contrast, charged vacancy defects were initialized from the fully optimized equilibrium geometries of their corresponding neutral states. Starting from these standardized initial configurations, independent structural relaxations were performed using both DFT and the MLIP. By comparing the atomic configuration and the total energy errors of the final converged structures, we systematically assessed the reliability of the MLIP as a surrogate for DFT in large-scale defect simulations.

\section{Results and Discussion}

\subsection{Performance for neutral defects}

\begin{figure*}[htbp]
    \centering
    \includegraphics[scale=0.7]{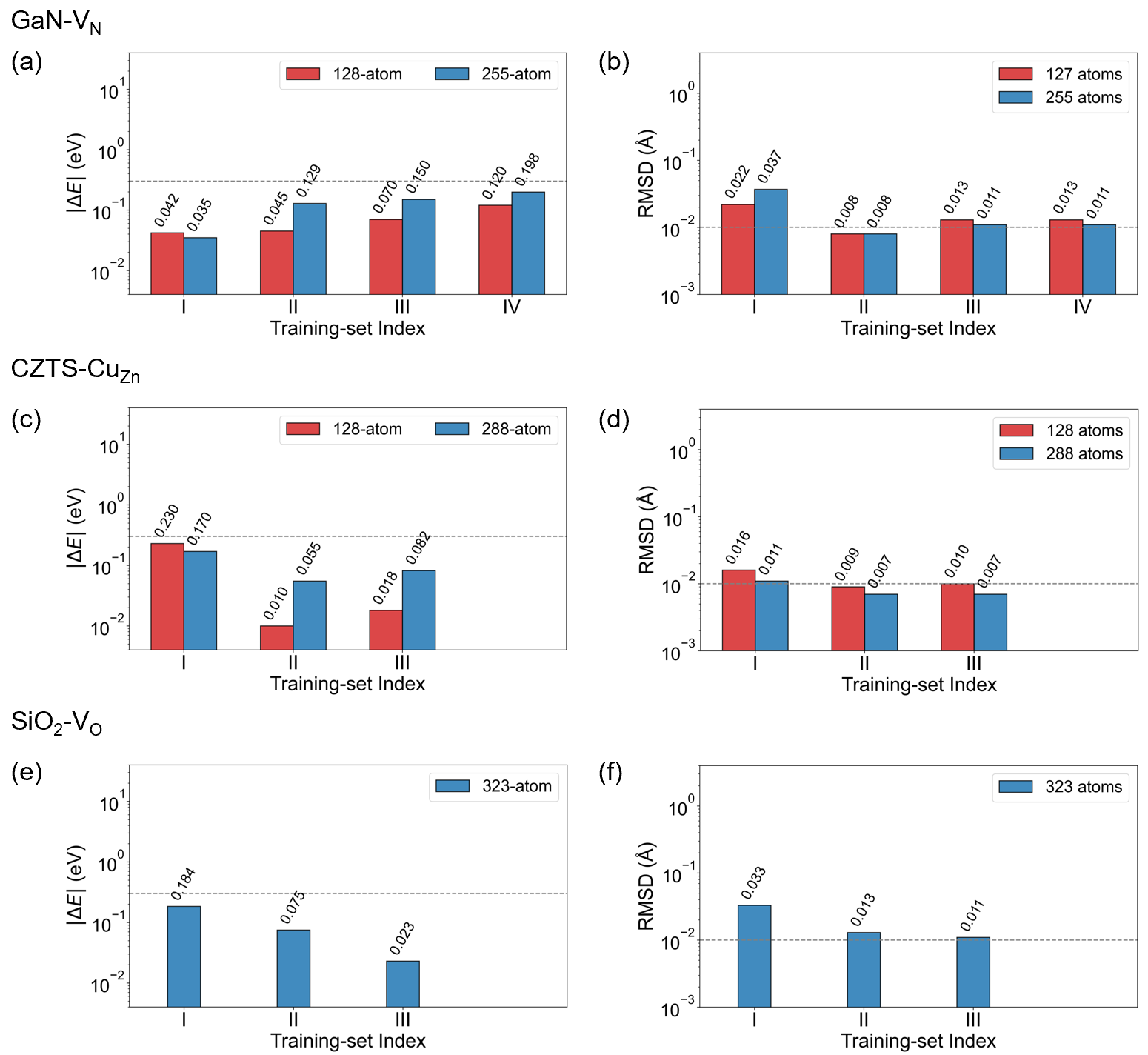}
    \caption{Total-energy prediction errors and RMSD of MLIP for neutral defect systems. (a)-(b) GaN$ -V_\mathrm{N}$,  (c)-(d) $\mathrm{Cu}_2\mathrm{ZnSnS}_4 - \mathrm{Cu}_\mathrm{Zn}$,  (e)-(f) SiO$_2 - V_\mathrm{O}$.}
    \label{fig:MLIP_neutral}
\end{figure*}

We first examine neutral defects as a reference case, where long-range electrostatic interactions associated with charged defects are absent. This test provides a baseline for evaluating whether standard MLIPs trained on limited small-supercell data can extrapolate to larger defect supercells. 
We evaluate the structural fidelity and energetic accuracy of MLIPs by calculating the root-mean-square displacement (RMSD) of atomic positions and total energy errors against DFT-relaxed reference structures. All corresponding results are summarized in Fig.~\ref{fig:MLIP_neutral}.

 Overall, the neutral-defect results indicate that standard MLIPs can provide stable structural predictions even when trained with small datasets. For GaN$ -V_{\mathrm{N}}$, the RMSD of the 255-atom test supercell decreases from 0.037~\AA{} for Training-set I to about 0.01~\AA{} after including additional small-supercell data. A similar trend is observed for SiO$_2 - V_{\mathrm{O}}$, where the RMSD of the 323-atom supercell decreases from 0.033~\AA{} to 0.011~\AA{}. For $\mathrm{Cu}_2\mathrm{ZnSnS}_4 - \mathrm{Cu}_{\mathrm{Zn}}$, the RMSD values are already small for Training-set I and remain below 0.02~\AA{} for all test supercells and training sets. These results show that the local structural relaxation around neutral defects can be well transferred from small to larger supercells.

The energy predictions also remain within a moderate error range for all neutral-defect systems. For $\mathrm{Cu}_2\mathrm{ZnSnS}_4 - \mathrm{Cu}_{\mathrm{Zn}}$, the energy error of the 128-atom test supercell decreases markedly after adding the 32-atom defect configuration, from 0.230~eV to 0.010~eV, while the error for the 288-atom test supercell remains below 0.17~eV for all training sets. For SiO$_2 - V_{\mathrm{O}}$, increasing the number of small-supercell defect configurations leads to a clear reduction of the energy error in the 323-atom test supercell, from 0.184~eV for Training-set I to 0.023~eV for Training-set III. In GaN$ -V_{\mathrm{N}}$, the energy errors show a less monotonic dependence on the training-set size and increase for the 255-atom test supercell when more small-supercell data are included. Nevertheless, all energy errors for neutral defects remain below 0.3~eV in the present tests.

These results suggest that, for neutral defects, the main local relaxation patterns can be learned from a small number of defect and host configurations, and the resulting standard MLIPs exhibit reasonable size-transferability to larger supercells. The structural predictions are robust, while the energy errors remain system dependent and do not always decrease monotonically with the amount of training data.

This neutral-defect benchmark serves as a useful reference for the charged-defect cases discussed below, where long-range electrostatic finite-size effects introduce additional difficulties for standard local MLIPs.

\subsection{ Performance for Charged Defects: The Defect-Only Scheme}

\begin{figure*}[htbp]
    \centering
    \includegraphics[scale=0.5]{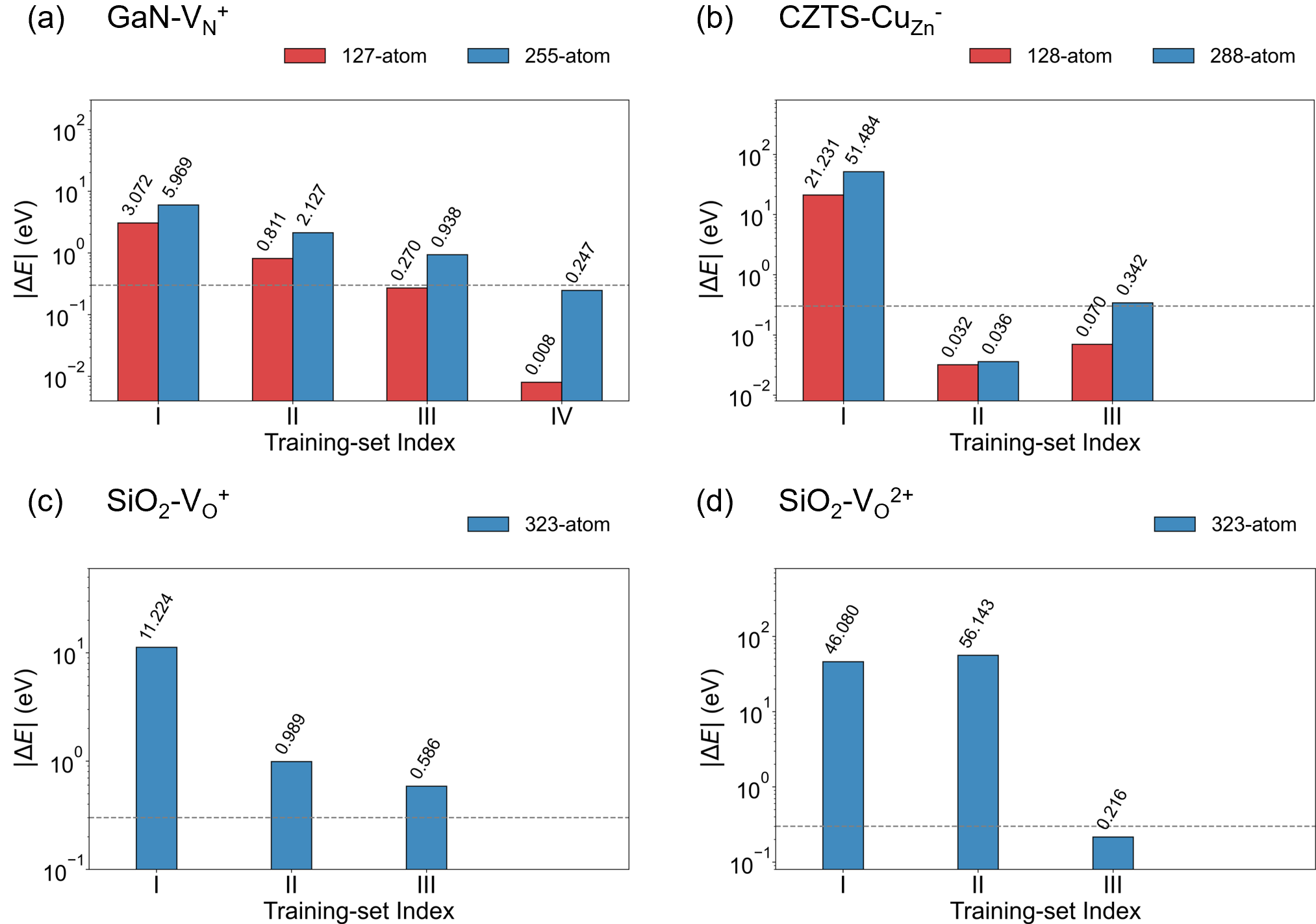}
    \caption{Total-energy prediction errors of MLIP trained using the Defect-Only Scheme in charged defect systems. (a) GaN$ -V_\mathrm{N}^{+}$, (b) $\mathrm{Cu}_2\mathrm{ZnSnS}_4 - \mathrm{Cu}_\mathrm{Zn}^{-}$, (c) SiO$_2 - V_\mathrm{O}^{+}$, and (d) SiO$_2 - V_\mathrm{O}^{2+}$.}
    \label{fig:MLIP_scheme1_energy}
\end{figure*}

Building upon the benchmark results for neutral defects, we proceed to investigate the model's capability in describing charged defect systems. 
We first consider the Defect-Only Scheme, in which only relaxed charged-defect configurations are used for training. The energy prediction errors of MLIP are shown in Fig.~\ref{fig:MLIP_scheme1_energy}. The results indicate that, when the training set contains charged-defect supercells of only a single size, MLIP is generally unable to extrapolate reliably to larger test supercells. Taking the $\mathrm{Cu}_2\mathrm{ZnSnS}_4 - \mathrm{Cu}_{\mathrm{Zn}}^{-}$ system as an example, when the training set consists only of 16-atom defect supercells, the total-energy prediction errors of the model for the 128-atom and 288-atom test supercells reach 21.231~eV and 51.484~eV, respectively. For the SiO$_2 - V_{\mathrm{O}}^{2+}$ system, when only 17-atom defect supercells are used for training, the error of the model for the 323-atom test supercell also reaches 46.080~eV. These results show that training only on small charged-defect configurations is insufficient to ensure reliable total-energy prediction by a standard local MLIP for larger charged-defect supercells.

After charged-defect configurations with multiple supercell sizes are introduced into the training set, the cross-size prediction capability of the model is generally improved substantially. For the GaN$ -V_{\mathrm{N}}^{+}$ system, when the training set is expanded from only 15-atom defect supercells to defect configurations containing 15, 31, 63, and 95 atoms, the total-energy prediction error of the model for the 127-atom test supercell decreases from 3.072~eV to 0.008~eV, and the error for the 255-atom test supercell also decreases from 5.969~eV to 0.247~eV. Similarly, for the SiO$_2 - V_{\mathrm{O}}^{2+}$ system, when the training set is further expanded to include 17-, 35-, and 71-atom defect configurations, the error of the model for the 323-atom test supercell eventually decreases to 0.216~eV.

\vspace{10pt}

Notably, the inclusion of multi-size training data does not guarantee a monotonic convergence of the energy error with respect to the larger supercell. For the SiO$_2 - V_{\mathrm{O}}^{2+}$ system, when the training set is expanded from only 17-atom defect configurations to both 17- and 35-atom defect configurations, the error of the model for the 323-atom test supercell instead increases from 46.080~eV to 56.143~eV; only after 71-atom defect configurations are further included does this error decrease to 0.216~eV. For the $\mathrm{Cu}_2\mathrm{ZnSnS}_4 - \mathrm{Cu}_{\mathrm{Zn}}^{-}$ system, when the training set is expanded from containing both 16- and 32-atom defect configurations to further including 64-atom defect configurations, the error of the model for the 288-atom test supercell also increases from 0.036~eV to 0.342~eV. In addition, for the SiO$_2 - V_{\mathrm{O}}^{+}$ system, although the error of the model for the 323-atom test supercell is significantly reduced from 11.224~eV after multi-size training data are added, the final error remains approximately 0.586~eV, which does not reach the error level observed in other systems with better convergence.

\begin{figure*}[htbp]
    \centering
    \includegraphics[scale=0.52]{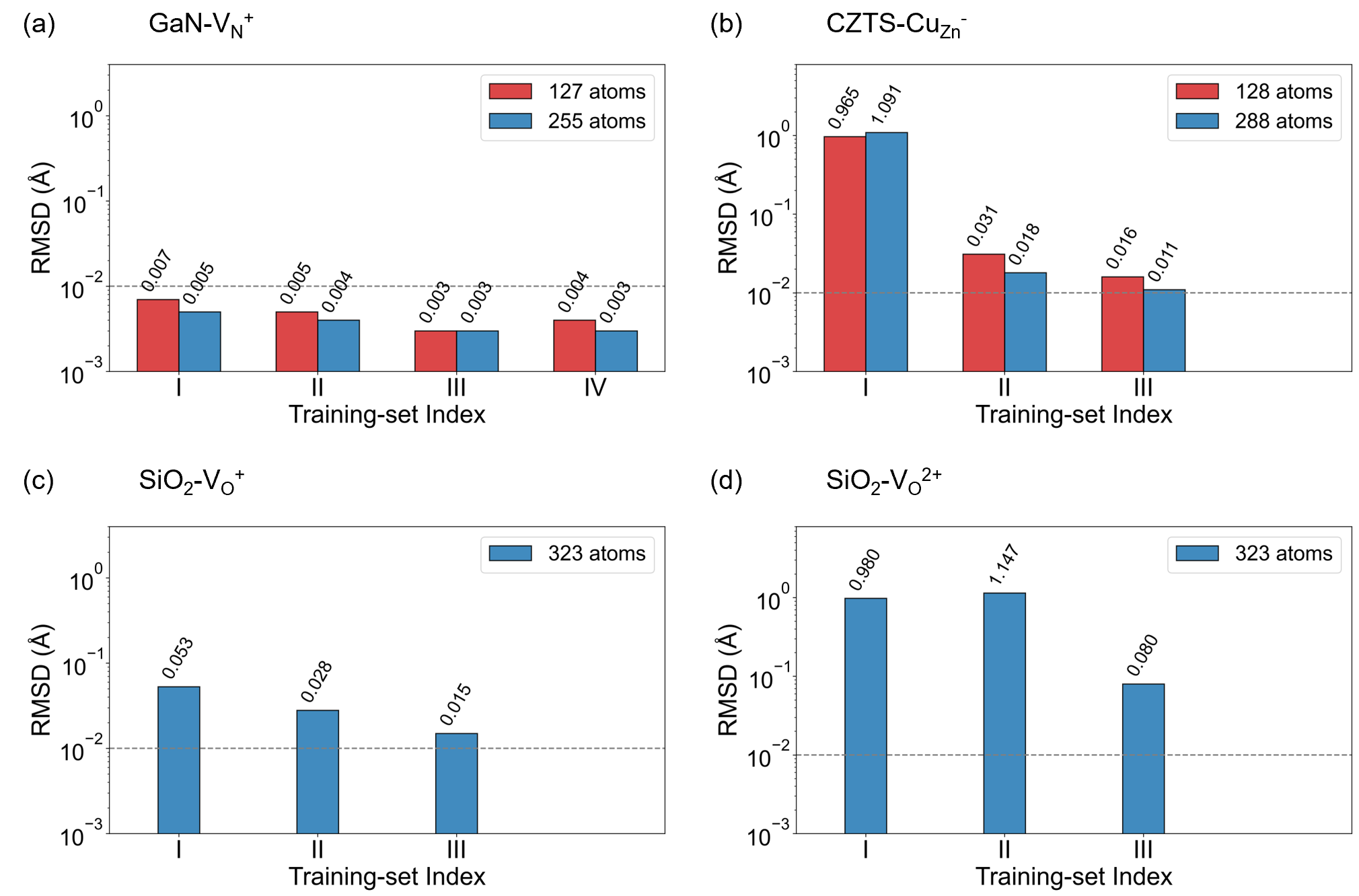}
    \caption{RMSD of MLIP-optimized structures trained using the Defect-Only Scheme relative to the reference structures in charged defect systems. (a) GaN$ -V_\mathrm{N}^{+}$, (b) $\mathrm{Cu}_2\mathrm{ZnSnS}_4 - \mathrm{Cu}_\mathrm{Zn}^{-}$, (c) SiO$_2 - V_\mathrm{O}^{+}$, and (d) SiO$_2 - V_\mathrm{O}^{2+}$. The dashed line indicates RMSD = 0.01~\AA{}.}
    \label{fig:MLIP_scheme1_rmsd}
\end{figure*}

Beyond energetic accuracy, we evaluate the structural fidelity by comparing the RMSD of MLIP-optimized geometries against DFT references (Fig.~\ref{fig:MLIP_scheme1_rmsd}). The results show that the description of structural relaxation by MLIP is also affected by the supercell-size composition of the training set, but its variation trend is not fully consistent with that of the total-energy error. For the GaN$ -V_{\mathrm{N}}^{+}$ system, even when only 15-atom charged-defect configurations are used for training, the RMSD values of the model for the 127-atom and 255-atom test supercells are only 0.007~\AA{} and 0.005~\AA{}, respectively. As 31-, 63-, and 95-atom defect configurations are further added to the training set, the RMSD remains at a low level of approximately 0.003--0.005~\AA{}. This indicates that, in this system, although the Defect-Only Scheme may produce relatively large total-energy prediction errors under single small-size training conditions, its prediction of the local relaxed structure remains relatively stable.

\vspace{5pt}

By contrast, the structural prediction results for the $\mathrm{Cu}_2\mathrm{ZnSnS}_4 - \mathrm{Cu}_{\mathrm{Zn}}^{-}$ system are more sensitive to the supercell-size composition of the training set. When the training set contains only 16-atom defect configurations, the RMSD values for the 128-atom and 288-atom test supercells reach 0.965~\AA{} and 1.091~\AA{}, respectively, indicating that a single small-size training set cannot reliably describe structural relaxation in large defect supercells. As 32-atom and 64-atom defect configurations are gradually added to the training set, the RMSD decreases significantly. Under the 16+32+64-atom training condition, the RMSD values of the model for the 128-atom and 288-atom test supercells decrease to 0.016~\AA{} and 0.011~\AA{}, respectively, approaching the level of 0.01~\AA{}.

The defect systems in SiO$_2$ show similar but more complex behavior. For the SiO$_2 - V_{\mathrm{O}}^{+}$ system, as larger defect configurations are introduced into the training set, the RMSD of the model for the 323-atom test supercell decreases from 0.053~\AA{} to 0.015~\AA{}, indicating that multi-size training can progressively improve the prediction of structural relaxation in this system. In contrast, the SiO$_2 - V_{\mathrm{O}}^{2+}$ system is more sensitive to the size composition of the training set. When the training set contains only 17-atom defect configurations, or contains both 17- and 35-atom defect configurations, the RMSD values of the model for the 287-atom test supercell reach 0.980~\AA{} and 1.147~\AA{}, respectively, indicating pronounced structural errors. Only after 71-atom defect configurations are further added does the RMSD decrease to 0.080~\AA{}.

Overall, the introduction of multi-size charged-defect configurations can not only improve total-energy prediction, but also help enhance the stability of structural relaxation prediction in some systems. However, the variations in structural errors and energy errors are not fully synchronized. For some charged defect systems, such as SiO$_2 - V_{\mathrm{O}}^{2+}$, introducing only small and medium-size defect configurations is still insufficient to constrain the structural relaxation behavior in large supercells, whereas the inclusion of larger training configurations may play a key role in improving structural prediction.

\subsection{ Performance for Charged Defects: the Defect+Host Scheme}

\begin{figure*}[htbp]
    \centering
    \includegraphics[scale=0.8]{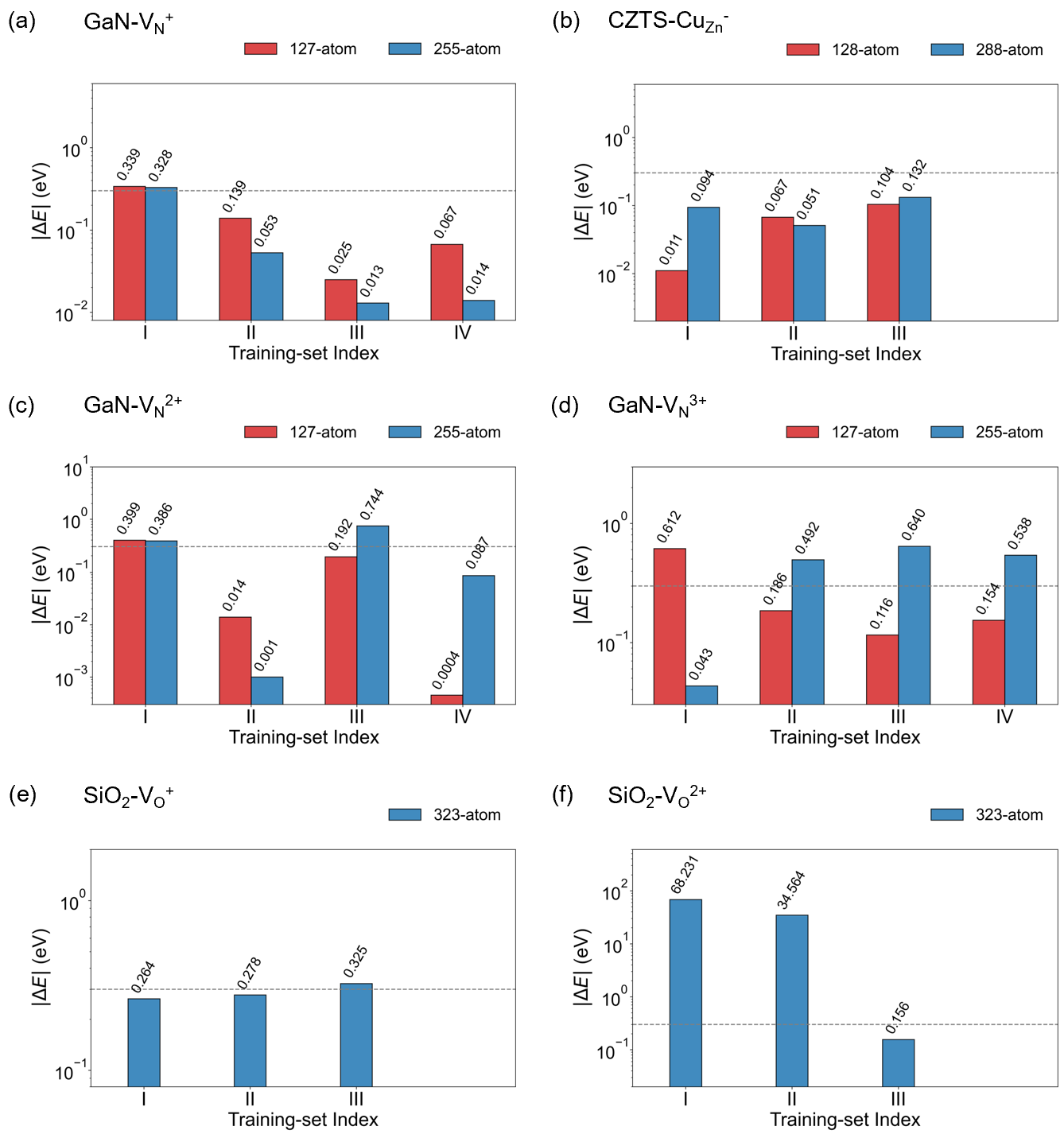}
    \caption{Total-energy prediction errors of MLIP trained using the Defect+Host Scheme in charged defect systems. (a) GaN$ -V_\mathrm{N}^{+}$, (b) $\mathrm{Cu}_2\mathrm{ZnSnS}_4 - \mathrm{Cu}_\mathrm{Zn}^{-}$, (c) GaN$ -V_\mathrm{N}^{2+}$, (d) GaN$ -V_\mathrm{N}^{3+}$, (e) SiO$_2 - V_\mathrm{O}^{+}$, and (f) SiO$_2 - V_\mathrm{O}^{2+}$.}
    \label{fig:MLIP_scheme2_energy}
\end{figure*}

To evaluate the impact of neutral bulk data on cross-size predictions, we next analyze the performance of the Defect+Host Scheme. The energy prediction errors under the Defect+Host Scheme are shown in Fig.~\ref{fig:MLIP_scheme2_energy}. The test results show that, compared with the Defect-Only Scheme, the inclusion of size-matched neutral bulk data significantly reduces the total-energy prediction errors for most large charged-defect systems. For the GaN$ -V_{\mathrm{N}}^{+}$ system, even when the training set contains only 15-atom charged-defect configurations, the energy errors of the Defect+Host Scheme for the 127-atom and 255-atom test supercells are reduced to 0.339~eV and 0.328~eV, respectively, which are much lower than those of the Defect-Only Scheme under the same conditions. As the training set is further expanded to include 15-, 31-, 63-, and 95-atom defect configurations, the errors of the Defect+Host Scheme for the two test supercells decrease to 0.067~eV and 0.014~eV, respectively, both remaining at low levels.

Similarly, the $\mathrm{Cu}_2\mathrm{ZnSnS}_4 - \mathrm{Cu}_{\mathrm{Zn}}^{-}$ system also shows clear improvement under the Defect+Host Scheme. When the training set contains only 16-atom charged-defect configurations, the errors of the Defect+Host Scheme for the 128-atom and 288-atom test supercells are $0.011$~eV and 0.094~eV, respectively, whereas the Defect-Only Scheme gives errors on the order of tens of eV under the same conditions. After 32-atom and 64-atom defect configurations are further introduced, the errors remain overall within approximately 0.1~eV. This indicates that, for the GaN$ -V_{\mathrm{N}}^{+}$ and $\mathrm{Cu}_2\mathrm{ZnSnS}_4 - \mathrm{Cu}_{\mathrm{Zn}}^{-}$ systems, size-matched neutral bulk data help provide an energy reference and reduce the errors caused by changes in supercell size.

\begin{figure*}[htbp]
    \centering
    \includegraphics[scale=0.78]{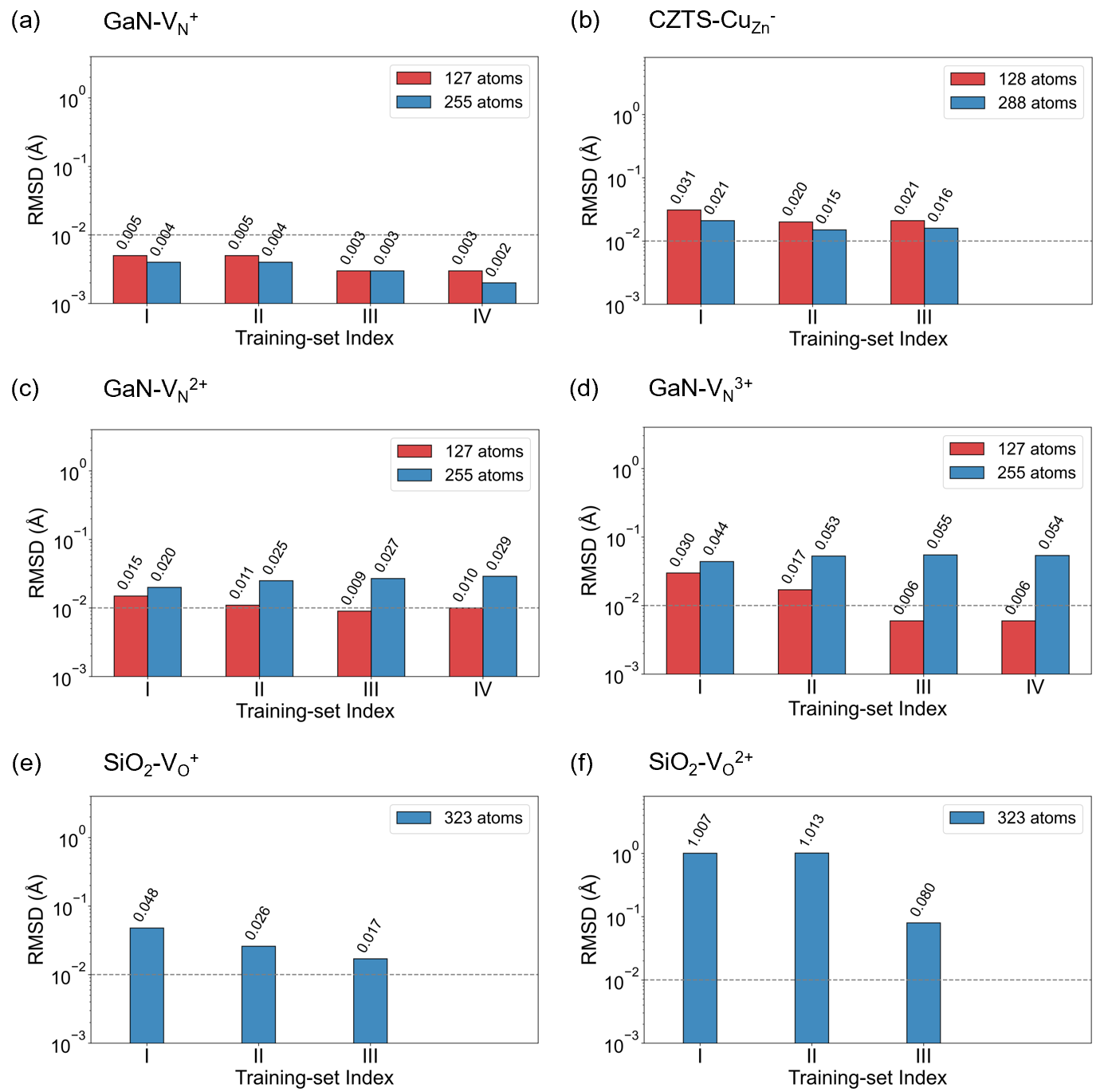}
    \caption{RMSD of MLIP-optimized structures trained using the Defect+Host Scheme relative to the reference structures in charged defect systems. (a) GaN$ -V_\mathrm{N}^{+}$, (b) $\mathrm{Cu}_2\mathrm{ZnSnS}_4 - \mathrm{Cu}_\mathrm{Zn}^{-}$, (c) GaN$ -V_\mathrm{N}^{2+}$, (d) GaN$ -V_\mathrm{N}^{3+}$, (e) SiO$_2 - V_\mathrm{O}^{+}$, and (f) SiO$_2 - V_\mathrm{O}^{2+}$.}
    \label{fig:MLIP_scheme2_rmsd}
\end{figure*}

For the SiO$_2 - V_\mathrm{O}^{+}$ system, the Defect+Host Scheme can also keep the energy errors for the large test supercell within a relatively small range. Under different training-set sizes, the errors of this system are mainly distributed around 0.3~eV with small overall fluctuations, indicating that, after neutral bulk data are included, the model predicts the energies of this charged defect state at different supercell sizes with relatively good stability.

However, the results for higher charge-state defects show that simply adding size-matched neutral bulk data is still insufficient to ensure stable cross-size prediction by the model when only small training sets are used. For the GaN$ -V_{\mathrm{N}}^{2+}$ system, when the training set contains only 15-atom defect configurations, the energy errors of the model for the 127-atom and 255-atom test supercells are 0.399~eV and 0.386~eV, respectively, both exceeding 0.3~eV. After 31-atom defect configurations are added, the errors for the two test supercells decrease to 0.014~eV and 0.001~eV, respectively. However, when 63-atom defect configurations are further included, the error for the 255-atom test supercell increases again to 0.744~eV. Only when the training set is expanded to include 15-, 31-, 63-, and 95-atom defect configurations does this error decrease again to 0.087~eV. For the GaN$ -V_{\mathrm{N}}^{3+}$ system, the energy error for the 127-atom test supercell can be reduced to below 0.154~eV after multi-size training data are introduced, but the error for the 255-atom test supercell remains in the range of 0.492--0.640~eV under most training configurations and does not show a stable decreasing trend as the training set is expanded.

This problem is more pronounced in the SiO$_2 - V_{\mathrm{O}}^{2+}$ system. When the training set contains only 17-atom defect configurations, or contains both 17- and 35-atom defect configurations, the energy errors still reach 68.231~eV and 34.564~eV, respectively. Only after 71-atom defect configurations are further added does the error decrease to 0.156~eV.

The structural relaxation results are broadly consistent with the above analysis of energy errors, and the corresponding structural errors are shown in Fig.~\ref{fig:MLIP_scheme2_rmsd}. For the GaN$ -V_{\mathrm{N}}^{+}$ system, the RMSD values obtained under the Defect+Host Scheme for different training-set sizes are all below 0.01~\AA{}, indicating that the MLIP-relaxed structures are very close to the DFT reference structures. The RMSD values of the $\mathrm{Cu}_2\mathrm{ZnSnS}_4 - \mathrm{Cu}_{\mathrm{Zn}}^{-}$ system under the Defect+Host Scheme are also much lower than the results obtained with the small training set in the Defect-Only Scheme, and remain stable in the range of approximately 0.015--0.031~\AA{}.

By contrast, the structural errors are more pronounced for higher charge-state defects. For the GaN$ -V_{\mathrm{N}}^{2+}$ system, the RMSD values for the 127-atom test supercell are 0.009--0.015~\AA{}, close to the level of 0.01~\AA{}. However, the RMSD values for the 255-atom test supercell are 0.020--0.029~\AA{}, clearly higher than those of the GaN$ -V_{\mathrm{N}}^{+}$ system. For the GaN$ -V_{\mathrm{N}}^{3+}$ system, the RMSD for the 127-atom test supercell decreases from 0.030~\AA{} to 0.006~\AA{} as the training set is expanded, whereas the RMSDs for the 255-atom test supercell are much higher, remaining in the range of 0.044--0.055~\AA{}. The SiO$_2 - V_{\mathrm{O}}^{2+}$ system exhibits a similar limitation. Under small training-set conditions, the RMSD remains around 1~\AA{}, and decreases to 0.080~\AA{} only after 71-atom charged-defect configurations are included, which is still larger than 0.01~\AA{}.

These findings demonstrate that while the Defect+Host Scheme effectively mitigates energy errors and structural deviations for low charge-state defects, its capability for higher charge-state defects still depends strongly on the size coverage of defect supercells in the training set.

\section{Conclusion}

In this work, we investigated the ability of standard MLIPs to predict the total energies and relaxed structures of charged defects across different supercell sizes. Using $V_{\mathrm{N}}$ in GaN, $V_{\mathrm{O}}$ in SiO$_2$, and $\mathrm{Cu}_{\mathrm{Zn}}$ in $\mathrm{Cu}_2\mathrm{ZnSnS}_4$ as representative defect systems, we proposed an efficient MLIP training scheme based on small supercells (less than 100 atoms) and limited DFT data. The training dataset can be constructed from only four DFT structural optimizations and subsequently used to predict charged-defect configurations in larger supercells.

Our results demonstrate that the composition and supercell size of the training data are crucial for the size extrapolation of standard MLIPs. Training exclusively on defect configurations from a single small supercell size leads to substantial errors in larger systems. Including defect configurations from multiple small supercell sizes improves both the energetic and structural predictions. The addition of size-matched neutral pristine host configurations further enhances the prediction accuracy for defects in low charge states. For highly charged defects, however, the finite-size effects associated with long-range electrostatic interactions remain significant, and small-supercell data alone are insufficient for reliable extrapolation. In such cases, larger charged-defect supercells must still be included in the training dataset. Overall, this work clarifies the practical capabilities and limitations of standard MLIPs for charged-defect simulations and provides a computationally efficient strategy for constructing reliable training datasets from limited DFT data. 

\begin{acknowledgments}

\end{acknowledgments}

\bibliography{apssamp}

\end{document}